\documentstyle[12pt]{article}
\title{If time is a local observable, then Hawking radiation is unitary}
\author{Hrvoje Nikoli\'c \\
Theoretical Physics Division, Rudjer Bo\v{s}kovi\'{c} Institute, \\
P.O.B. 180, HR-10002 Zagreb, Croatia \\
{\normalsize e-mail: hnikolic@irb.hr} \\
\makebox[1in]{} \\
}
\date{\today}
\begin{document}
\maketitle
\begin{abstract}
In the usual formulation of quantum theory, time is a global classical evolution parameter, not a
local quantum observable. On the other hand, both canonical quantum gravity (which lacks
fundamental time-evolution parameter) and the principle of spacetime covariance (which
insists that time should be treated on an equal footing with space) suggest that quantum
theory should be slightly reformulated, in a manner that promotes time to a local
observable. Such a reformulated quantum theory is unitary in a more general sense than the
usual quantum theory. In particular, this promotes the non-unitary Hawking radiation to a
unitary phenomenon, which avoids the black-hole information paradox.
\end{abstract}
\vspace*{0.5cm}

\section{Introduction}

Black-hole information paradox \cite{hawk2,gid,har,pres,pag,gid2,str,math,hoss,fabbri} 
is one of the most controversial conceptual puzzles in modern physics.
Recently, a very active debate on the paradox has been stimulated by the work of
Almheiri, Marolf, Polchinski and Sully \cite{AMPS}. In this work, they start from 
3 reasonable assumptions
\begin{enumerate}
\item Hawking radiation is unitary,
\item low energy field theory is valid near horizon, 
\item freely falling observer sees nothing special at the horizon,
\end{enumerate}
and derive a contradiction.
To resolve the contradiction, they argue that assumption 3. is wrong, i.e.
that there must be a firewall at the horizon seen even by freely falling observers.
In this paper we defend a less popular possibility, that it is assumption 1. which is wrong.
In other words, we defend the possibility that Hawking radiation is {\em not} unitary.

How can the absence of unitarity be compatible with quantum mechanics (QM)?
We argue that Hawking radiation is not ``unitary'' 
in the usual meaning of this word, but is unitary 
in a slightly {\em generalized} sense.
In this way all 3 assumptions can be simultaneously satisfied, 
if only the notion of ``unitarity'' is slightly generalized. 

The main idea rests on the fact that
the usual notion of ``unitarity'' means
that the {\em time evolution} is unitary.
In our proposal of generalized unitarity
there is time, but there is no time {\em evolution}. 
Therefore, without time-evolution in general, 
there can be no non-unitary time evolution in particular.
But if there is no time evolution, then what time is?
Our answer is that time is a {\it local observable}.
By contrast, in standard QM
time is neither local nor an observable.
It is not an observable 
because in a time-dependent state $|\psi(t)\rangle$, 
$t$ is an external classical parameter, not a quantum operator.
It is not local 
because there is only one parameter $t$, 
which parametrizes the {\em whole} space-like hypersurface 
on which $|\psi\rangle$ is defined.

One way to generalize time $t$ to a local quantity is by the
Tomonaga-Schwinger formalism \cite{tomonaga,schw}, in which one makes the replacement
\begin{equation}
 t \rightarrow T({\bf x}) ,
\end{equation}
so that each space-point ${\bf x}$ has another time parameter $T({\bf x})$.
In this way, the Schr\"{o}dinger equation
\begin{equation}
 H\psi(t)=i\frac{\partial}{\partial t}\psi(t)
\end{equation}
generalizes to the Tomonaga-Schwinger equation
\begin{equation}
 {\cal H}({\bf x}) \psi[T] = i \frac{\delta}{\delta T({\bf x})} \psi[T] ,
\end{equation}
where ${\cal H}({\bf x})$ is the local Hamiltonian density.
But in this formalism, $T({\bf x})$ is still a collection of (infinitely many) 
classical parameters, not a collection of quantum observables.
Quantum state is still defined on an (arbitrarily curved) 
space-like hypersurface with fixed $T({\bf x})$.

So, how to make $t$ a quantum observable?
For that purpose consider the quantum probability density
\begin{equation}
 p(q,t)=|\psi(q,t)|^2 .
\end{equation}
When we say that $t$ is a classical parameter, we mean that $p(q,t)$ 
is the probability of $q$ {\it at} $t$, i.e. that probability obeys 
\begin{equation}
 \int dq \, p(q,t)= 1 \;\;\; {\rm for \;\; all \;\;} t . 
\end{equation}
This property corresponds to unitarity in the usual sense.

Analogously, to say that $t$ is a quantum observable, means that $p(q,t)$ 
is the probability of $q$ {\it and of} $t$, i.e. that probability obeys
\begin{equation}
 \int dq \, dt \, p(q,t)= 1  .
\end{equation}
This property corresponds to our unitarity in the generalized sense.

Such a generalized unitarity implies that $t$ is just like any other observable $q$.
For example, since evolution of space (for constant time) does not make sense, 
it implies that evolution of time itself also does not make sense. 
This suggests the block-universe picture of the world, according to which
there is no evolution of time at the fundamental level, while
past, presence and future all ``simultaneously'' exist.

The block-universe picture has both advantages and disadvantages.
The main advantage is a consistency with classical relativity,
because $t$ is treated on an equal footing with ${\bf x}$ 
and spacetime is viewed as a single 4-dimensional object.
The main disadvantage is a
contradiction with our intuitive (psychological) experience of time. 
In this paper we study what we can cure 
if we try to swallow the counter-intuitive pill  
of the block universe. In particular, we study how it helps 
to solve the black-hole information paradox.

Our main result can be understood in simple terms as follows.
For times after the evaporation, the inside particle 
does not longer exist, which implies that 
information about that particle is destroyed.
But if we think of it as a block universe,
the inside particle is not really destroyed;  
it exists in the past!

For such an interpretation to work, however, it is important that time is {\em local}.
Namely, locality implies that each particle has its own time variable,  
so different particles may co-exist at {\em different} times.

For such an interpretation to work, it is also important that time is an {\em observable}.
Namely, it guarantees that ``co-existence'' at different {\em times} 
should be interpreted in the same way as co-existence 
at different positions in {\em space}. 

In the rest of the paper it remains to see how to explicitly realize this general idea 
in a more concrete theoretical framework.
We discuss two different approaches. In Sec.~\ref{SEC2} we study
canonical quantum gravity, while in Sec.~\ref{SEC3}
we study a space-time covariant theory. The conclusions are drawn in Sec.~\ref{SEC4}.

\section{Canonical approach}
\label{SEC2}

\subsection{Canonical quantum gravity and the concept of time}

Canonical quantum gravity is based on the Hamiltonian constraint
\begin{equation}\label{eq1}
 {\cal H}\Psi[g,\phi] =0 ,
\end{equation}
where $\cal{H}$ is the Hamiltonian-density operator and $\Psi[g,\phi]$ is the
wave function of the universe, depending on gravitational and matter degrees
of freedom denoted by $g$ and $\phi$, respectively. 
(On the technical level,
the most promising variant of (\ref{eq1}) is based on loop quantum gravity
\cite{rov}, where $g$ denotes the loop variables.) 
Clearly, $\Psi[g,\phi]$ does not depend on an external time parameter,
which is often referred to as problem of time in quantum gravity
(see e.g. \cite{isham,kuchar} for older reviews and \cite{rov}
for a review written from a more modern perspective).
Obviously, since $\Psi[g,\phi]$ does not depend on time, the information
encoded in $\Psi[g,\phi]$ cannot depend on time either, i.e. 
information cannot be ``lost''. 
The lack of time dependence can be thought of as ``time evolution''
described by a trivial unitary operator
\begin{equation}\label{eq2}
 U(t)\equiv 1 ,
\end{equation}
which means that the theory is unitary in a trivial sense.
The quantity
\begin{equation}
 \rho[g,\phi]=\Psi^*[g,\phi] \Psi[g,\phi]  
\end{equation}
can be interpreted as probability of given values $g$ and $\phi$, provided
that $\Psi[g,\phi]$ is normalized such that
\begin{equation}
 \int {\cal D}g \, {\cal D}\phi \, \Psi^*[g,\phi] \Psi[g,\phi] =1.
\end{equation}
In loop quantum gravity, the formal measure ${\cal D}g$ 
is mathematically well defined, and there are justified expectations
that ${\cal D}\phi$ could be well defined too.

Even though there is no fundamental notion of time, a phenomenological
notion of time can still be introduced. The most physical way to do it is to
introduce a clock-time \cite{rov}. Essentially, this means that some of the
matter degrees of freedom describe the reading of a ``clock''. In this case the
Hamiltonian $H=\int d^3x \, {\cal H}$ can be split as
\begin{equation}
H=\tilde{H} +H_{\rm clock} ,
\end{equation}
where $H_{\rm clock}$ describes the clock and $\tilde{H}$ is the rest of the Hamiltonian.
The Hamiltonian for a good clock can be approximated by a Hamiltonian of the form
\begin{equation}\label{H_clock}
 H_{\rm clock} \simeq \lambda P_{\rm clock} ,
\end{equation}
where $\phi\equiv \{\tilde{\phi},Q_{\rm clock}\}$, 
$Q_{\rm clock}$ is the configuration variable representing the reading
of the clock, $P_{\rm clock}$ is the canonical momentum conjugated to $Q_{\rm clock}$,
and $\lambda$ is a coupling constant.
Indeed, the resulting classical equation of motion
\begin{equation}
 \frac{dQ_{\rm clock}}{dt}= \frac{\partial H_{\rm clock}}{\partial P_{\rm clock}} \simeq \lambda
\end{equation}
implies
\begin{equation}
 Q_{\rm clock}(t) \simeq \lambda t ,
\end{equation}
so $Q_{\rm clock}$ increases approximately linearly with time, which means that the value of 
$Q_{\rm clock}$ is a good measure of time. 

In quantum theory the momentum
$P_{\rm clock}$ is the derivative operator
\begin{equation}
 P_{\rm clock} = -i\frac{\partial}{\partial Q_{\rm clock}},
\end{equation}
so (\ref{H_clock}) can be written as
\begin{equation}
 H_{\rm clock} \simeq -i\frac{\partial}{\partial q_{\rm clock}} ,
\end{equation}
where $q_{\rm clock}\equiv \lambda^{-1} Q_{\rm clock}$.
In this way, (\ref{eq1}) implies a Schrodinger-like equation
\begin{equation}\label{sch}
 \tilde{H}\Psi[g,\tilde{\phi},q_{\rm clock}] \simeq i\frac{\partial}{\partial q_{\rm clock}} 
\Psi[g,\tilde{\phi},q_{\rm clock}] .
\end{equation}

Even though (\ref{sch}) has the same form as the usual Schr\"odinger equation,
we stress two important differences with respect to the
usual interpretation of time in the Schr\"odinger equation. 
First, $q_{\rm clock}$ is a {\em quantum} observable,
not a classical external parameter. Second, in most cases $q_{\rm clock}$
is a {\em local} quantity, not a quantity that can be associated
with a whole spacelike hypersurface. As we shall see, these two features 
are essential for our resolution of the black-hole information paradox.

%Finally note that even though the analysis above involves matter, the existence of matter
%is in no way essential. In particular, in the absence of matter the role of a ``clock''
%can be taken by one of the gravitational degrees of freedom.
%(Recall that gravity, just as most other field theories, contains an infinite
%number of the degrees of freedom.)
%Therefore, even though in the next section we discuss information paradox
%in the presence of matter, essentially the same analysis can be applied
%in the absence of matter as well.

\subsection{Implications on black-hole information paradox}

Now assume that $\Psi[g,\phi]$ is a solution of (\ref{eq1}) that describes
an evaporating black hole. Of course, an explicit construction of such a
solution is prohibitively difficult. Yet, under reasonable assumptions justified
by understanding of semi-classical black holes,
some qualitative features of such a hypothetical solution
can easily be guessed without an explicit solution at hand. In particular, 
it is reasonable to assume
that, at least approximately, the degrees of freedom can be split into inside and outside
degrees of freedom.
Therefore we write
\begin{equation}\label{wf}
\Psi[g,\phi] = \Psi[g_{\rm in},\phi_{\rm in},g_{\rm out},\phi_{\rm out}] .
\end{equation}
This state can also be represented by a pure-state density matrix
\begin{equation}\label{pure}
 \rho[g_{\rm in},\phi_{\rm in},g_{\rm out},\phi_{\rm out}|
g'_{\rm in},\phi'_{\rm in},g'_{\rm out},\phi'_{\rm out}] =
\Psi[g_{\rm in},\phi_{\rm in},g_{\rm out},\phi_{\rm out}]
\Psi^*[g'_{\rm in},\phi'_{\rm in},g'_{\rm out},\phi'_{\rm out}] .
\end{equation}
By tracing out over the inside degrees of freedom,
we get the mixed-state density matrix
\begin{equation}
 \rho_{\rm out}[g_{\rm out},\phi_{\rm out}|g'_{\rm out},\phi'_{\rm out}] =
\int {\cal D}g_{\rm in} \, {\cal D}\phi_{\rm in} \,
\rho[g_{\rm in},\phi_{\rm in},g_{\rm out},\phi_{\rm out}|
g_{\rm in},\phi_{\rm in},g'_{\rm out},\phi'_{\rm out}] ,
\end{equation}
which describes information available to an outside observer.
Next we identify a clock-time of an outside observer, so that
we can write
\begin{equation}
\rho_{\rm out}[g_{\rm out},\phi_{\rm out}|g'_{\rm out},\phi'_{\rm out}] =
\rho_{\rm out}[g_{\rm out},\tilde{\phi}_{\rm out},q_{\rm clock \; out}|
g'_{\rm out},\tilde{\phi}'_{\rm out},q'_{\rm clock \; out}] .
\end{equation}
Finally, by considering the clock-diagonal matrix elements 
$q_{\rm clock \; out}=q'_{\rm clock \; out}\equiv t$, we get an ``evolving''
outside density matrix
\begin{equation}\label{t-evol}
\rho_{\rm out}[g_{\rm out},\tilde{\phi}_{\rm out}|
g'_{\rm out},\tilde{\phi}'_{\rm out}](t) \equiv
\rho_{\rm out}[g_{\rm out},\tilde{\phi}_{\rm out},t|
g'_{\rm out},\tilde{\phi}'_{\rm out},t] . 
\end{equation}
Clearly, the $t$-evolution described by (\ref{t-evol}) may not be unitary.
At times $t$ for which the black hole has evaporated completely, 
(\ref{t-evol}) may correspond to a mixed state, in accordance with 
predictions of the semi-classical theory \cite{hawk2}.
One could think that it is merely a restatement of the information paradox,
but it is actually much more than that. Unlike the standard statement of the paradox
\cite{hawk2}, such a restatement contains also a resolution of the paradox.
Namely,
from the construction of (\ref{t-evol}) it is evident that there is nothing
fundamental about such a violation of unitarity. No information is really lost.
The full information content is encoded in the 
pure state (\ref{pure}) equivalent to the wave function (\ref{wf}).
This is very different from the information loss in the standard formulation
\cite{hawk2}, where information seems to be really lost and no description
in terms of pure states seems possible.

To see more explicitly where the information is hidden, it is useful
to introduce {\em two} clocks, such that (\ref{wf}) can be written as
\begin{equation}\label{wf2}
\Psi[g_{\rm in},\phi_{\rm in},g_{\rm out},\phi_{\rm out}] =
\Psi[g_{\rm in},\tilde{\phi}_{\rm in},q_{\rm clock \; in},
g_{\rm out},\tilde{\phi}_{\rm out},q_{\rm clock \; out}].
\end{equation}
Here $q_{\rm clock \; in}$ and $q_{\rm clock \; out}$ are configuration
variables describing an inside clock and an outside clock, respectively.
Assuming that the black hole eventually evaporates completely,
the inside clock cannot show a time larger than some value $t_{\rm evap}$
corresponding to the time needed for the complete evaporation.
More precisely, the probability that $q_{\rm clock \; in}>t_{\rm evap}$
is vanishing, so 
\begin{equation}\label{wf3}
 \Psi[g_{\rm in},\tilde{\phi}_{\rm in},q_{\rm clock \; in},
g_{\rm out},\tilde{\phi}_{\rm out},q_{\rm clock \; out}]  =0 \;\;\;
{\rm for} \;\;\; q_{\rm clock \; in}>t_{\rm evap} .
\end{equation}
The existence of the wave function (\ref{wf2}) implies that 
the system can be described by a pure state even after the complete evaporation.
However, this description is trivial, because (\ref{wf3}) says that the wave function
has a vanishing value
for $q_{\rm clock \; in}>t_{\rm evap}$. Still, even a nontrivial pure-state description for 
$q_{\rm clock \; out}>t_{\rm evap}$ is possible, provided that $q_{\rm clock \; in}$ 
is restricted to the
region $q_{\rm clock \; in}<t_{\rm evap}$. In this case (\ref{wf2}) describes the
correlations between the outside degrees of freedom after the complete evaporation and 
the inside degrees of freedom before the complete evaporation. 
In other words, if one asks where the 
information after the complete evaporation is hidden, then the answer is -- 
{\em it is hidden in the past}.
Of course, experimentalists cannot travel to the past, so information is lost for the experimentalists.
Yet, this information loss is described by a pure state, so one does not need to use the
Hawking formalism \cite{hawk2} in which a state evolves from a pure to a mixed state.
By avoiding this formalism one avoids its pathologies \cite{banks} too, which may be 
viewed as the main advantage of our approach. 

One might object that information hidden in the past is the same as 
information destruction, but it is not.
The difference is subtle and essential for our approach, so let us explain 
it once again more carefully. Information hidden in the past and information destruction
are the same for an observer who views the world as an entity
that evolves with time $t$ in (\ref{t-evol}). However, such a view of the world
is emergent rather than fundamental, because time is emergent
rather than fundamental. At the fundamental level there is no time
and no evolution. The fundamental world is static and unitary, as described by
(\ref{eq2}). 
The concept of ``past'' refers to something which does not longer exist at the
emergent level, but it still exists at the fundamental level.
Thus, at the fundamental level, information
is better described as being {\em present} in the past and only {\em hidden}
for an emergent observer, rather than being destroyed.
In this sense, our resolution of the information paradox does not remove
the non-unitary time evolution entirely. Instead,  
{\em it shifts the non-unitary time evolution from a fundamental level to an emergent one}.

One might still argue that we have only shifted the problem (from one level
to another) and not really solved it. But in our view such a shift of the problem
is also a solution, or at least a crucial part of a solution.
Namely, it is typical for emergent theories in physics that they lack full self-consistency,
even when the underlying fundamental theories are self-consistent.
Indeed, a presence of an inconsistency in an otherwise successful physical theory 
is often a sign that this theory is not fundamental, but 
emergent. (A classic example is the ultraviolet catastrophe in classical statistical
mechanics. It was resolved by Planck and others by recognizing that 
classical statistical mechanics emerges from 
more fundamental quantum statistical mechanics, which does not involve
the ultraviolet catastrophe. In this way the inconsistency of classical statistical
mechanics was not removed, but shifted from a fundamental to an emergent
level.)
In our case of the black-hole information paradox, the emergent theory is not self-consistent 
as it violates unitarity. 
%
%(In other words, the time evolution of the system is described by the Hawking non-unitary
%operator $\$ $ \cite{hawk2}, which leads to inconsistencies \cite{banks}.)
%
We resolve the problem by identifying a more fundamental unitary theory
from which the unitarity-violating theory emerges. 
The unitarity violation is nothing but a sign 
that {\em the emergent description in terms of time evolution is not fully applicable
to the phenomenon of black-hole evaporation,
and the fundamental theory involving no time evolution is a more appropriate
description.}
In our opinion, it is a legitimate resolution
of the black-hole information paradox, even if an unexpected one. 

\section{Covariant approach}
\label{SEC3}

The covariant approach has already been presented elsewhere \cite{nik,nik2}
(see also \cite{hartle1,hartle2}),
so here we only sketch the main ideas.
For simplicity, we consider a system with a fixed number of particles, but 
the generalization to uncertain number of particles is also possible \cite{nik}.

\subsection{Many-time formalism and spacetime probability}

A single-time wave function in standard QM
\begin{equation}
 \psi({\bf x}_1,\ldots ,{\bf x}_n,t) 
\end{equation}
generalizes to the many-time wave function 
\cite{dirac,rosenfeld,dfp,bloch,tomonaga,tumulka}
\begin{equation}
\psi({\bf x}_1,t_1\ldots ,{\bf x}_n,t_n) . 
\end{equation}
The standard single-time wave function is just a special case of the many-time wave function
\begin{equation}
 \psi({\bf x}_1,\ldots ,{\bf x}_n,t) =
\psi({\bf x}_1,t_1\ldots ,{\bf x}_n,t_n) |_{t_1=\cdots =t_n=t} .
\end{equation}
The many-time wave function allows a covariant notation
\begin{equation}
\psi({\bf x}_1,t_1\ldots ,{\bf x}_n,t_n) = \psi(x_1, \ldots ,x_n) ,
\end{equation}
where $x=({\bf x}, t)$.

The probability $dP$ that particles will be found around points $x_1, \ldots ,x_n$ is postulated to be
\cite{nikijqi09,nikijmpa,nikijqi11}
\begin{equation}\label{4-prob}
 dP=|\psi(x_1, \ldots ,x_n)|^2 d^4x_1\cdots d^4x_n .
\end{equation}
This is the probability in {\em spacetime}, which was first proposed in \cite{stuec}.

Concerning the $n$-particle interpretation of the wave function $\psi(x_1, \ldots ,x_n)$,
there is a conceptual subtlety which requires a clarification. The $n$-point wave function
$\psi(x_1, \ldots ,x_n)$ is obtained by acting $n$ times with the particle-creation operator 
on the vacuum \cite{nikijmpa}.
In this sense, even when the separation between the points $x_1,\ldots, x_n$ is timelike,  
these points should be interpreted as positions of $n$ separate particles, not as 
one particle measured at $n$ times. However, in curved spacetime the choice of the vacuum
(which determines the corresponding particle-creation operators) is not unique \cite{birdel}.
Different splits of spacetime into space and time lead to different ``natural'' choices
of the vacuum and the corresponding particles. A way to choose the vacuum and particles in a unique manner 
is to introduce a preferred time, which seems particularly plausible in the context of Horava gravity 
\cite{nik_horava}.

From (\ref{4-prob}), the standard probability in {\em space} can be recovered in the following way. 
If one {\em knows} that particles are detected at times 
\begin{equation}
t_1=\cdots =t_n=t ,
\end{equation}
then space probability is the {\em conditional probability}  
\begin{equation}
dP_{(3n)}=\frac{|\psi({\bf x}_1,\ldots,{\bf x}_n,t)|^2 d^3x_1 \cdots d^3x_n}
{N_{t}} ,
\end{equation}
with the normalization factor
\begin{equation}
 N_{t}=\int |\psi({\bf x}_1,\ldots,{\bf x}_n,t)|^2 d^3x_1 \cdots d^3x_n .
\end{equation}
This is the standard probability in space.

\subsection{Application to black-hole information paradox}

Assume that black hole evaporates completely after time $t_{\rm evap}$.  
This implies that the probability of detecting inside particle for $t>t_{\rm evap}$ is zero, so 
the inside modes satisfy
\begin{equation}
 \psi^{\rm(in)}_l({\bf x}_{\rm in},t_{\rm in}) |_{t_{\rm in}>t_{\rm evap}}= 0 .
\end{equation}
The outside modes $\psi^{\rm(out)}_k(x_{\rm out})$ do not vanish for $t>t_{\rm evap}$. 

Now consider a Hawking pair of particles. 
Their most general entangled wave function is
\begin{equation}
 \psi(x_{\rm in}, x_{\rm out})=\sum_{l}\sum_{k}c_{lk}\psi^{\rm(in)}_l(x_{\rm in}) \psi^{\rm(out)}_k(x_{\rm out}) ,
\end{equation}
which is equivalent to the pure-state density matrix
\begin{equation}
 \rho(x_{\rm in}, x_{\rm out}|x'_{\rm in}, x'_{\rm out})=
\psi(x_{\rm in}, x_{\rm out}) \psi^*(x'_{\rm in}, x'_{\rm out}) .
\end{equation}
The outside state is a mixed state
\begin{equation}
 \rho^{\rm(out)}(x_{\rm out}|x'_{\rm out})=\int d^4x_{\rm in}\, |g(x_{\rm in})|^{1/2} \,
\rho(x_{\rm in}, x_{\rm out}|x_{\rm in}, x'_{\rm out}) .
\end{equation}

Now the information paradox resolves in a way similar to that 
in the canonical approach. 
First, time is {\em local} because each particle at space-position ${\bf x}_a$ 
has its own time $t_a$. In general $t_1 \neq t_2 \neq \cdots \neq t_n$. 
Second, time is an {\em observable} because 
in (\ref{4-prob}) time
is interpreted in the same way 
as the space position.
Therefore, different particles may co-exist at different times. 
As a consequence information is not lost, in the sense that the wave function describes  
a correlation of the outside particle after the evaporation 
with the inside particle before the evaporation.

\section{Conclusion}
\label{SEC4}

The possibility that information is destroyed after the black-hole evaporation 
is not ruled out.
In this paper we have argued that this destruction may be an illusion, in the sense that information 
may still be hidden in the past. Such a possibility has both advantages and disadvantages.
A disadvantage is the fact
that such a view contradicts our intuitive notion of time evolution.
An advantage is the fact that it looks natural from the mathematical point of view. 

In particular, such a view is natural in the canonical approach. 
Namely, canonical quantum gravity lacks a fundamental notion
of time evolution, which implies trivial unitarity of the theory at the fundamental level.
Time and evolution are emergent concepts, defined with the aid of a physical clock.
In general, such a clock-time only has a local meaning and is represented 
by a quantum observable. 

In addition, such a view is also natural in the covariant approach, if we insist that time should be treated 
on an equal footing with space.

In both approaches, 
information present on a global
spacelike hypersurface does not play any fundamental role. Consequently,
even if observers living after the complete evaporation of a black hole
cannot see all information encoded in the wave function of the universe,
which can be interpreted as effective violation of unitarity for the observers, 
the full wave function of the universe still contains all the information
and no fundamental violation of unitarity takes place.
In this way both fundamental unitarity and phenomenological 
information loss may peacefully coexist,
which resolves the black-hole information paradox.

In this way, both approaches support the idea that if time is a local observable, 
then Hawking radiation is unitary.

\section*{Acknowledgements}

This work was supported by the Ministry of Science of the
Republic of Croatia under Contract No.~098-0982930-2864.

\end{document}